\documentclass[conference]{IEEEtran}

\IEEEoverridecommandlockouts

\usepackage{cite}
\usepackage{amsmath}
\usepackage{amssymb}
\usepackage{graphicx}
\usepackage{textcomp}
\usepackage{xcolor}
\usepackage{flushend}
\usepackage{lipsum}
\def\BibTeX{{\rm B\kern-.05em{\sc i\kern-.025em b}\kern-.08em
T\kern-.1667em\lower.7ex\hbox{E}\kern-.125emX}}
\usepackage{algorithm}
\usepackage{algpseudocode}
\usepackage{soul}
\renewcommand{\baselinestretch}{0.99}
\begin{document}
\title{Joint Phase Shift Optimization and Precoder Selection for RIS-Assisted 5G NR MIMO Systems}

\author{\IEEEauthorblockN{Osman Mert Yilmaz\IEEEauthorrefmark{1}\IEEEauthorrefmark{4}, Tayfun Yilmaz\IEEEauthorrefmark{2}\IEEEauthorrefmark{3}\IEEEauthorrefmark{4},  Ali Gorcin\IEEEauthorrefmark{1}\IEEEauthorrefmark{4}, Ibrahim Hokelek\IEEEauthorrefmark{4}, Ertugrul Guvenkaya\IEEEauthorrefmark{1}, Haci Ilhan\IEEEauthorrefmark{3} }

\IEEEauthorblockA{\IEEEauthorrefmark{1} Department of Electronics and Communication Engineering, Istanbul Technical University, {\.{I}}stanbul, Türkiye} 

\IEEEauthorblockA{\IEEEauthorrefmark{2} Department of Aviation Electrics and Electronics, Kocaeli University, {Kocaeli, Türkiye} }

\IEEEauthorblockA{\IEEEauthorrefmark{3} Department of Electronics and Communication Engineering, Yildiz Technical University, {\.{I}}stanbul, Türkiye}

\IEEEauthorblockA{\IEEEauthorrefmark{4}  Communications and Signal Processing Research (HİSAR) Lab., T{\"{U}}B{\.{I}}TAK B{\.{I}}LGEM, Kocaeli, Türkiye}

\IEEEauthorblockA{Emails: \it yilmazo24@itu.edu.tr, \it
    tayfun.yilmaz@kocaeli.edu.tr, \it
    aligorcin@itu.edu.tr, \\ \it  
       ibrahim.hokelek@tubitak.gov.tr, eguvenkaya@itu.edu.tr, \it
     ilhanh@yildiz.edu.tr, \it }
}

\maketitle
\begin{abstract}
By intelligently reconfiguring wireless propagation environment, reconfigurable intelligent surfaces (RISs) can enhance signal quality, suppress interference, and improve channel conditions, thereby serving as a powerful complement to multiple-input multiple-output (MIMO) architectures. However, jointly optimizing the RIS phase shifts and the MIMO transmit precoder in 5G and beyond networks remains largely unexplored. This paper addresses this gap by proposing a singular value ($\lambda$)-based RIS optimization strategy, where the phase shifts are configured to maximize the dominant singular values of the cascaded channel matrix, and the corresponding singular vectors are utilized for MIMO transmit precoding. The proposed precoder selection does not require mutual information computation across subbands, thereby reducing time complexity. To solve the $\lambda$-based optimization problem, maximum cross-swapping algorithm (MCA) is applied while an effective rank-based method is utilized for benchmarking purposes. The simulation results show that the proposed precoder selection method consistently outperforms the conventional approach under $\lambda$-based RIS optimization. 

\end{abstract}


\section{Introduction} 
 The proliferation of internet of things (IoT) and smart city applications along with emerging ultra-low latency services such as autonomous driving, has exposed the limitations of 5G in terms of data rate, latency, coverage, energy efficiency, and mobility. Academics driven system-oriented research and innovation
activities have been focusing on the development of 6G \cite{survey6gwireless}, where higher
frequencies with a large amount of spectrum resources are considered towards satisfying the enormous data rate requirements. However, 6G brings new challenges in terms of energy consumption, deployment and operation costs, and exacerbated propagation losses due to higher frequency bands~\cite{IRSAWCTutorial}. To overcome these challenges, the beamforming technology using a large antenna array such as reconfigurable intelligent surfaces (RISs) are envisioned as key enablers in 6G, thanks to their ability to dynamically adjust element phases for signal steering and loss mitigation~\cite{RISfor6g, guo2024deeplearningbasedcsifeedback}. Nevertheless, technical challenges associated to a joint optimization of the RIS phase shifts and the MIMO precoding need to be studied.

Recent studies on RIS-assisted MIMO (a.k.a MIMO-RIS) networks have gained momentum to assess the feasibility of integrating RIS technology into 5G NR MIMO systems. The communication performance in MIMO systems can be significantly improved through a joint optimization of the precoding vector at the gNodeB and the phase shifts of the RIS elements \cite{wu2019intelligentreflectingsurfaceenhanced,guo2019weightedsumratemaximizationreconfigurable}. In MIMO systems of cellular wireless networks such as LTE, LTE-Advanced, and 5G NR, the user equipment (UE) estimates the wireless channel using pilot signals and reports channel state information (CSI) to the gNodeB in a compressed form standardized by 3rd Generation Partnership Project (3GPP) as it would require a large amount of uplink (UL) resources if they were transmitted without any compression. The compressed CSI includes precoding matrix indicator (PMI), rank indicator (RI), and channel quality index (CQI). PMI represents the optimal precoding matrix based on channel conditions as determined by the UE. 3GPP has standardized two codebooks for 5G NR, namely Type-I and Type-II Codebooks \cite{etsi2020138214}. There are studies providing a comprehensive analysis of 3GPP codebooks including the structures of codebooks, precoding matrix selection, and the optimization of CSI feedback \cite{jin2022massivemimoevolution3gpp,tutdlprecoderselection,qin2023reviewcodebookscsifeedback}. However, to the best of the authors' knowledge, there is no study that jointly optimizes the phase shifts of RIS elements and the transmit precoder of a MIMO system in such a way that backwards compatibility with existing 3GPP standards is ensured. Furthermore, an additional complexity of the RIS phase sift optimization prioritizes the development of a low-complexity precoder selection method by utilizing a singular value decomposition (SVD) based technique. In light of these considerations, the main contributions of the paper can be summarized as

\begin{itemize}
    \item  This paper proposes a joint optimization of the RIS phase shifts and the MIMO precoder, where the phase shifts of the RIS elements are optimized by maximizing the singular values obtained from the SVD of the cascaded MIMO channel matrix, and subsequently, the corresponding singular vectors are utilized for the MIMO precoder selection. The proposed method specifically designed for a MIMO system is fully consistent with 5G NR standards.
    \item The proposed precoder selection strategy is independent of the number of MIMO layers during the computation of the wideband part, and does not require mutual information calculation for the subband part. As a result, the time complexity is reduced, especially for the scenarios involving a large number of antennas and subbands.
    \item An effective rank-based optimization criterion is adopted from the literature as a benchmark to the singular value ($\lambda$)-based approach, where the optimal RIS phase shift values are obtained using the maximum cross-swapping algorithm (MCA).
    \item A realistic RIS channel model is developed within the MATLAB 5G Toolbox for the performance evaluation. The simulation results demonstrate that the proposed precoder selection method outperforms conventional schemes under both $\lambda$- and effective rank-based optimization strategies, particularly for the scenarios with low to moderate numbers of antennas and RIS elements. Additionally, $\lambda$-based optimization consistently yields higher SNR performance than effective rank-based optimization.
\end{itemize}
\section{System Model}
\begin{figure*}[htbp]
\centerline{\includegraphics [scale=.7]{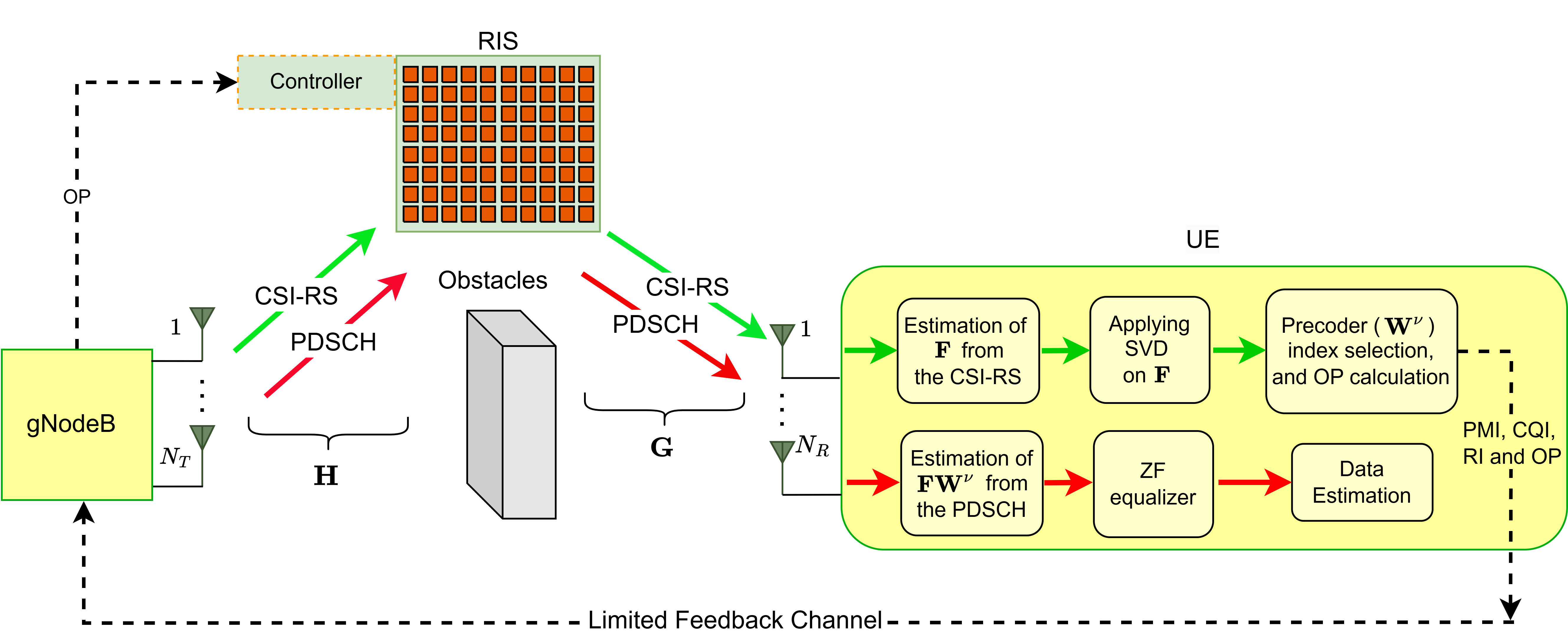}}
\caption{Illustration of the proposed RIS-assisted MIMO system for 5G NR.}
\label{sys_mod}
\end{figure*}

Fig.~\ref{sys_mod} illustrates an RIS-assisted 5G-NR MIMO system, where a gNodeB with $N_T$ transmit antennas communicates with a UE equipped with $N_R$ receive antennas. Due to the obstacles, no direct link exists between the gNodeB and the UE. An RIS with $N_{\text{RIS}} = N_{\text{RIS}}^x \times N_{\text{RIS}}^y$-reflecting elements serves as a passive relay to establish an alternative communication link consisting of cascaded channels. The gNodeB employs a cross-polarized planar array, where each polarization uses $N_T/2$ physical antennas. Through precoding, these are mapped to $P_{\text{CSI-RS}}$ antenna ports arranged in a two-dimensional antenna array with $N_1$ and $N_2$ ports in the horizontal and vertical directions, respectively, such that $P_{\text{CSI-RS}} = 2N_1N_2$. For analytical simplicity, we assume $P_{\text{CSI-RS}} = N_T$. The gNodeB transmits channel state information reference signal (CSI-RS) through the RIS, and the UE estimates the channel accordingly. The cascaded RIS-assisted MIMO channels between the gNodeB and the UE can be modeled as
\begin{equation}
    \mathbf{F} = \mathbf{G} \mathbf{\Phi} \mathbf{H},\label{eq}
\end{equation}
where $\mathbf{F} \in \mathbb{C}^{N_{\text{R}} \times P_{\text{CSI-RS}}}$. Here, $\mathbf{H} \in \mathbb{C}^{N_{\text{RIS}} \times P_{\text{CSI-RS}}}$ represents the channel matrix between the gNodeB and the RIS while $\mathbf{G} \in \mathbb{C}^{N_R \times N_{\text{RIS}}}$  describes the channel matrix between the RIS and the UE\footnote{\textit{Notations:} Throughout this paper, bold lowercase and uppercase letters represent vectors and matrices, respectively.}. The entries of these matrices are assumed to follow the clustered delay line (CDL) channel models. The delay profile has been selected based on the CDL-C specified in \cite{etsi2020138901}, Section 7.7.1, Table 7.7.1-3. $\mathbf{\Phi} \in \mathbb{C}^{N_{\text{RIS}} \times N_{\text{RIS}}}$ is a diagonal matrix composed of the reflection coefficients of each unit cell. The reflection matrix $\mathbf{\Phi}$ can be represented as 
\begin{equation}
    \mathbf{\Phi} = \text{diag} \left\{ A_{1}e^{j\theta_1}, A_{2}e^{j\theta_2}, \dots, A_{N_{\text{RIS}}}e^{j\theta_{N_{\text{RIS}}}} \right\},
\end{equation}
where $A_n$ and $e^{j\theta_n}$ denote the amplitude response and phase shift of the unit cell, respectively, with $\theta_n \in [-\pi, \pi)$ and $A_n \in [0,1]$. This study examines a scenario in which the reflection phase is chosen from a finite set of discrete values based on a defined quantization level. So the phase configuration vector as $\boldsymbol{\theta} = [\theta_1, \theta_2, \dots, \theta_{N_{\text{RIS}}}]$,  
where each phase $\theta_n$ is constrained to belong to a predefined discrete set, i.e., $\forall \theta_n \in \Xi_b = \left\{ 0, \frac{2\pi}{2^b}, \dots, \frac{2\pi(2^b -1)}{2^b} \right\}$.  
In this context, $b$ denotes the quantization bit width, which determines the granularity of the discrete phase levels for each unit cell. For the simplicity and brevity, the amplitude response parameter $A_n$ is assumed to be identical for all elements.

After estimating the cascaded channel matrix of $\mathbf{F}$, the next step as shown in Fig. \ref{sys_mod} is to apply SVD on the estimated $\mathbf{F}$ to obtain its singular values and vectors. A parameter to be used for the RIS phase optimization (i.e., optimization parameter - OP) is calculated using these singular values. Finally, the precoder from the Type-I codebook is selected using the singular vectors corresponding to the highest OP for the RIS. Then, the CSI reports and OP are fed back to the gNodeB through a limited feedback channel to optimize the RIS phase shifts and the MIMO precoding at the gNodeB. The precoded data is generated using the selected precoder and transmitted by the gNodeB towards the UE.Once the signal is received by the UE through the physical downlink shared channel (PDSCH), the precoded channel matrix information $\mathbf{FW}^{\nu}$, with $\mathbf{W}^{\nu}$ denoting the $\nu$ layer precoding matrix, is estimated from the precoded data. After applying a zero forcing (ZF) equalizer, the data is obtained in the UE side. The following sections provide mathematical foundations for joint optimization of phase shifts and precoding in MIMO-RIS systems.

\subsection{RIS-Assisted MIMO System with Linear Precoding}\label{AA}

In the baseband representation of a MIMO-RIS wireless communication system, the received signal per resource element (RE), corresponding to each orthogonal frequency division multiplexing (OFDM) symbol and subcarrier, can be formulated as follows:
\begin{equation}
    \mathbf{y} = \sqrt{\dfrac{1}{\text{PL}_{\text{RIS}}}} \mathbf{F} \mathbf{W}^{\nu} \mathbf{x} + \mathbf{n},
\end{equation}
where $\nu \geq 1$ depicts the number of layers, and $\mathbf{W}^\nu \in \mathbb{C}^{P_{\text{CSI-RS}} \times \nu}$ denotes the precoding matrix for $\nu$ layer MIMO system. gNodeB selects the precoding matrix $\mathbf{W}^\nu$ from a predefined codebook based on the CSI report transmitted by UE over the limited feedback channel. Additionally, we incorporated the OP into the CSI report in the proposed system. $\mathbf{x} \in \mathbb{C}^{\nu \times 1}$ is the transmitted signal vector, satisfying $ \mathbb{E} \left[ \mathbf{x} \mathbf{x}^{H} \right] = \dfrac{P_{T}}{N_{T}}\mathbf{I}_{\nu}$, where $\text{P}_{T}$ is the transmit power. $\mathbf{n} \in \mathbb{C}^{N_R \times 1}$ is the additive white Gaussian noise (AWGN) vector with $\mathbb{E} \left[ \mathbf{n}\mathbf{n}^H \right] = \sigma^2 \mathbf{I}_{N_R},$ where $\sigma^2$ denotes the noise variance. $\text{PL}_{\text{RIS}}$ represents path loss for a RIS scenario in the far field case and defined in \cite{etsigr2023003111} as
\begin{equation}
    \text{PL}_{\text{RIS}} = \frac{64\pi^3 (d_t d_r)^2}{G_t G_r G_s N_{\text{RIS}}^2 d_x d_y z^2 F(\alpha_t, \beta_t) F(\alpha_r, \beta_r) A_{n}^2},
\end{equation}
where \( G_s = \frac{4\pi d_x d_y}{z^2} \) represents the scattering gain of an individual RIS element, while \( G_t \) and \( G_r \) denote the transmit and receive antenna gains. The distances from the gNodeB to the RIS and from the RIS to the UE are denoted by \( d_t \) and \( d_r \). The physical dimensions of each RIS element are given by \( d_x \) and \( d_y \), representing its width and length. The wavelength is denoted by \( z \). The normalized radiation patterns in the transmission and reception directions are given by \( F(\alpha_t, \beta_t) \) and \( F(\alpha_r, \beta_r) \). The angles \( \alpha_t \) and \( \beta_t \) define the elevation and azimuth from the RIS center to the transmitter. As stated in~\cite{plris}, \( F(\alpha, \beta) \) can be expressed as
\begin{equation}
    F(\alpha, \beta) =
    \begin{cases}
    \cos^3 \alpha, & \alpha \in \left[ 0, \frac{\pi}{2} \right], \beta \in [0, 2\pi] \\
    0, & \alpha \in \left( \frac{\pi}{2}, \pi \right], \beta \in [0, 2\pi]
    \end{cases},
\end{equation}
As depicted in Fig. 1, we apply the ZF equalizer in the receiver, where the ZF equalization matrix is
\begin{equation}
    \begin{aligned}
        \mathbf{W}_{\text{zf}} 
        & = \sqrt{\text{PL}_{\text{RIS}}}\left((\mathbf{W}^\nu)^H \mathbf{F}^H \mathbf{F} \mathbf{W}^\nu\right)^{-1}(\mathbf{W}^\nu)^H \mathbf{F}^H,
    \end{aligned}
\end{equation}
where $\left(.\right)^{H}$ denotes the conjugate transpose (hermitian) operation for matrix. By left-multiplying the received signal vector $\mathbf{y}$ by $\mathbf{W}_{\text{zf}}$, the received signal is obtained as
\begin{equation}
\label{eqn:ZF_EQ}
    \begin{aligned}
        \hat{\mathbf{x}} & = \mathbf{W}_{\text{zf}} \mathbf{y} = \left((\mathbf{W}^\nu)^H \mathbf{F}^H \mathbf{F} \mathbf{W}^\nu\right)^{-1} (\mathbf{W}^\nu)^H \mathbf{F}^H\mathbf{F} \mathbf{W}^\nu \mathbf{x} \\
        & \qquad + \sqrt{{\text{PL}_{\text{RIS}}}}\left((\mathbf{W}^\nu)^H \mathbf{F}^H \mathbf{F} \mathbf{W}^\nu\right)^{-1}(\mathbf{W}^\nu)^H \mathbf{F}^H \mathbf{n}.
    \end{aligned}
\end{equation}
Consequently, using (\ref{eqn:ZF_EQ}), the instantaneous received SNR at the output of the ZF receiver for the \( r \)-th layer can be expressed as
\begin{equation}
\label{ZF-SNR}
\rho_{r}^{\text{ZF}} = \frac{\bar{\rho}}{\left[\left(\mathbf{W}^\nu\right)^H \mathbf{F}^H \mathbf{F} \mathbf{W}^\nu\right]_{r}^{-1}},
\end{equation}
where \([\,\cdot\,]_{r}\) denotes the \( r \)-th diagonal element of the matrix and \( r \in \{1, \dots, \nu\} \). Here, \( \bar{\rho} \) represents the average SNR which is calculated as
$\bar{\rho} = \frac{P_{T}}{\sigma^2 N_{T} \text{PL}_{\text{RIS}}}$.
Based on (\ref{ZF-SNR}) and $\left[\text{\citenum{riserank}, Eq.~(12)}\right]$, the achievable rate at the UE can be expressed as
\begin{equation}
\label{achv_rate}
C^{\nu} = \sum_{r=1}^{\nu} \log_{2} \left( 1 + \bar{\rho} \left[\left(\mathbf{W}^\nu\right)^H \mathbf{F}^H \mathbf{F} \mathbf{W}^\nu\right]_{r} \right),
\end{equation}

In 5G NR MIMO systems, only PMI is fed back to the gNodeB to limit the control signaling overhead. Without such constraints, the $r$-th singular vector from the SVD of the cascaded channel matrix could be directly transmitted. However, due to the limited feedback channel, the system cannot directly transmit the singular vectors. Instead, the PMI can be determined by extracting the singular vectors from the SVD. The proposed precoder selection method, leveraging both the dominant and second-largest singular vectors, is detailed in Section II-C2.
\subsection{3GPP Type-I Codebook Structure for 5G NR MIMO Systems}
 This study considers the Single Panel Type-I codebook which takes into account the Inverse Discrete Fourier Transform (IDFT)-based grid of beams in their precoder structures. Each IDFT beam \( \mathbf{v}_{l,m} \in \mathbb{C}^{N_1 N_2 \times 1} \) in the grid of beams is constructed using the Kronecker product of a horizontal beam \( \mathbf{v}'_l \) and a vertical beam \( \mathbf{u}_m \). This is formulated as follows
\begin{subequations}
    \begin{equation} \label{eq:codebook}
        \mathbf{v}_{l,m} = \mathbf{v}'_l \otimes \mathbf{u}_m
    \end{equation} 
    \begin{equation}
        \mathbf{v}'_l = \begin{bmatrix} 
        1 & e^{\frac{j2\pi l}{O_1 N_1}} & \dots & e^{\frac{j2\pi l (N_1-1)}{O_1 N_1}} 
        \end{bmatrix}^T_{1 \times N_1}
    \end{equation}  
    \begin{equation}
        \mathbf{u}_m = 
        \begin{bmatrix} 
        1 & e^{\frac{j2\pi m}{O_2 N_2}} & \dots & e^{\frac{j2\pi m (N_2-1)}{O_2 N_2}} 
        \end{bmatrix}^T_{1 \times N_2}
    \end{equation}
\end{subequations}
where $O_1$ and $O_2$ are the oversampling factors of the horizontal and vertical dimensions, respectively, and $l \in \{0,1, \dots, N_1 O_1 -1\} \quad \text{and} \quad m \in \{0,1, \dots, N_2 O_2 -1\}$.

The Type-I codebook \( \mathbf{W}^\nu \) for layer-\( \nu \) is constructed as the product of two matrices, \( \mathbf{W}_1^\nu \) and \( \mathbf{W}_2^\nu \), which correspond to wideband and subband precoding, respectively

\begin{equation}
   \mathbf{W}^{\nu} = \frac{1}{\sqrt{\nu \cdot P_{\text{CSI-RS}}}} \mathbf{W}_1^{\nu} \mathbf{W}_2^{\nu}.
\end{equation}

For layer-1 $\mathbf{W}_1^1 \in \mathbb{C}^{2N_1N_2 \times 2}$ is a block diagonal wideband beam matrix containing the IDFT vector $\mathbf{v}_{l,m}$ on the main diagonal. The IDFT vector $\mathbf{v}_{l,m}$ represents a beam selected from the grid of beams and remains unchanged across $N_3$ subbands. The matrix \( \mathbf{W}_2^1 \) is specified for each \( t \)-th subband in the layer-1 case, where the phase entry \( \varphi_{n_t} \) is defined as \( \varphi_{n_t} = e^{\frac{j\pi n_t}{2}} \) and selected from the discrete set \( \{1, j, -1, -j\} \). In this case, \( \mathbf{W}_2^1 \) can be represented by a co-phasing vector \( \mathbf{w}_2 \in \mathbb{C}^{2} \). The wideband precoder \( \mathbf{W}_1^1 \) and the co-phasing vector \( \mathbf{w}_2 \) are given by
    \begin{equation*}
        \mathbf{W}_1^1 = \begin{bmatrix}
        \mathbf{v}_{l,m} & \mathbf{0}_{N_1 N_2 \times 1} \\
        \mathbf{0}_{N_1 N_2 \times 1} & \mathbf{v}_{l,m}
        \end{bmatrix},
      \text{and}~\mathbf{w}_2 = \dfrac{1}{\sqrt{P_{\text{CSI-RS}}}}
        \begin{bmatrix}
        1 \\
        \varphi_{n}
        \end{bmatrix}.
    \end{equation*}

For the layer-1 case, the precoder structure corresponding to the \( t \)-th subband (\( t = 0, \dots, N_3 - 1 \)) is defined as
\begin{equation}
    \begin{aligned}
        \mathbf{W}_t^1
        &= \frac{1}{\sqrt{P_{\text{CSI-RS}}}} \mathbf{W}_1^1 \mathbf{w}_2
        = \frac{1}{\sqrt{P_{\text{CSI-RS}}}} 
        \begin{bmatrix} 
        \mathbf{v}_{l,m} \\ 
        \varphi_{n_t} \mathbf{v}_{l,m} 
        \end{bmatrix}.
    \end{aligned}
\end{equation}

For layer-2 and layer-3,4 $\left(P_{\text{CSI-RS}} < 16\right)$ transmission, the matrix $\mathbf{W}_1^{\nu}\ \in \mathbb{C}^{2N_1N_2 \times 4}$, where $\nu \in \{2, 3, 4\}$, can be generalized as
\begin{equation} \label{eq:W1}
    \mathbf{W}_1^{\nu} = \begin{bmatrix}
    \mathbf{v}_{l,m} & \mathbf{v}_{l',m'} & \mathbf{0}_{N_1 N_2 \times 1} & \mathbf{0}_{N_1 N_2 \times 1} \\
    \mathbf{0}_{N_1 N_2 \times 1} & \mathbf{0}_{N_1 N_2 \times 1} & \mathbf{v}_{l,m} & \mathbf{v}_{l',m'}
    \end{bmatrix}
\end{equation} 
where $\mathbf{v}_{l',m'}$ denotes a possibly distinct beam selected from the grid of beams according to the rules specified in Table 5.2.2.2.1-3 and Table 5.2.2.2.1-4 of Release-16 \cite{etsi2020138214}. The co-phasing value $\varphi_{n_t}$ is selected from the discrete set $\{1, j\}$ in transmission layer-2,3,4. However, the structure of $\mathbf{W}_2^{\nu}$ varies for each of these transmission layers\footnote{The matrices \( \mathbf{W}_2^{\nu} \) corresponding to layer-2,3, and 4 in the Type-I codebook structure can be found in~\cite[Eq. (15)--(17)]{tutdlprecoderselection}. For brevity, the explicit matrix forms are omitted in this paper.}.

\subsection{Precoder Selection Methods for RIS-assisted 5G NR MIMO Systems}
This section presents the proposed precoder selection method for RIS-assisted 5G NR MIMO systems when an SVD-based RIS optimization is applied. Furthermore, key differences and similarities between the proposed algorithm and the conventional precoder selection strategy\footnote{For a comprehensive overview of the conventional precoder selection method, the readers are referred to \cite{tutdlprecoderselection}.} are highlighted, and their time complexities are analyzed.

\subsubsection{Proposed precoder selection method}
The method enables standard PMI feedback to the base station by mapping the information obtained from SVD to 3GPP codebooks. The SVD of the estimated cascaded MIMO channel matrix $\mathbf{F}$, which is derived from CSI-RS measurements, is expressed as follows
\begin{equation}
    \mathbf{F} = \mathbf{U} \mathbf{\Sigma} \mathbf{V}^{H},
\end{equation}
where $\mathbf{\Sigma} \in \mathbb{C}^{N_{\text{R}} \times P_{\text{CSI-RS}}}$ matrix having in its diagonal the singular values of $\mathbf{F}$ $\left( \text{i.e.,} \mathbf{\Sigma} = \text{diag}(\lambda_1, \dots, \lambda_{\text{min}\left(N_{\text{R}},P_{\text{CSI-RS}}\right)}) \right)$, $\mathbf{U} \in \mathbb{C}^{N_{\text{R}} \times N_{\text{R}}}$ unitary matrix having as columns the left-singular vectors of $\mathbf{F}$, $\mathbf{V} \in \mathbb{C}^{P_{\text{CSI-RS}} \times P_{\text{CSI-RS}}}$ unitary matrix having as columns the right-singular vectors of $\mathbf{F}$. Optimal precoders $\mathbf{w}_{\text{opt}} \ \text{and} \ \mathbf{w}_{\text{opt}_\text{sub}}\in\mathbb{C}^{P_{\text{CSI-RS}}}$ are defined as dominant and the second-largest singular vectors.
 
The first step of the proposed precoder selection method, the optimal beam \( \mathbf{v}_{l^*,m^*} \) is assumed to be identical for both polarizations and is selected as follows
\begin{equation} \label{eq:pro_first}
    \mathbf{v}_{l^*,m^*} = \arg\max_{\mathbf{v}_{l,m} \in \mathcal{B}} \mathbf{v}_{l,m}^{H} \mathbf{w}^*_{\text{opt}},
\end{equation}
where \( \mathcal{B} \) is the IDFT grid of beams (The parameters $N_1$, $N_2$, $O_1$, and $O_2$ are configured according to \eqref{eq:codebook}.), and $\mathbf{w}^*_{\text{opt}} \in \mathbb{C}^{P_{\text{CSI-RS}}/2}$ is the first half of $\mathbf{w}_{\text{opt}}$. In scenarios where \( \nu > 1 \), the optimal beam \( \mathbf{v}_{l^{*'},m^{*'}} \) must also be selected. Although a similar inner product–based criterion is employed, it is applied using the first half of the second largest singular vector of \( \mathbf{F} \), denoted by \( \mathbf{w}^*_{\text{opt}_{\text{sub}}} \), and a distinct beam set \( \mathcal{B'} \). The beam selection is given by
\begin{equation} \label{eq:pro_second}
    \mathbf{v}_{l^{*'},m^{*'}} = \arg\max_{\mathbf{v}_{l',m'} \in \mathcal{B'}} \mathbf{v}_{l',m'}^{H} \mathbf{w}^*_{\text{opt}_{\text{sub}}},
\end{equation}
where $\mathbf{w}^*_{\text{opt}_{\text{sub}}} \in \mathbb{C}^{P_{\text{CSI-RS}}/2}$ is the first half of $\mathbf{w}_{\text{opt}_{\text{sub}}}$. Here, $\mathcal{B'}$ denotes the beam grid constructed according to the rules specified in Tables 5.2.2.2.1-3 and 5.2.2.2.1-4 of 3GPP Release-16~\cite{etsi2020138214}, and the number of beams in $\mathcal{B'}$, denoted by $N_{\mathcal{B'}}$, is generally small.

In the second step, to determine the co-phasing phase $\varphi_{n_t}$ for each subband $t$, the co-phasing index $n_t$ is obtained by performing the following operation
\begin{equation}
\label{eqn:co-phase}
    {n_t} = \underset{n}{\mathrm{argmax}}
    \begin{bmatrix} 
       \mathbf{v}_{l^*,m^*} \\ 
       \varphi_{{n}} \mathbf{v}_{l^*,m^*}
   \end{bmatrix}^{H} 
   \mathbf{w}_{\text{opt}},
\end{equation}
where $n$ is selected from the set $\{0,1,2,3\}$ when $\nu = 1$, and from the set $\{0,1\}$ when $\nu > 1$.

\begin{algorithm}[b]
\caption{Proposed Precoder Selection Method} \label{alg:precoder}
\begin{algorithmic}[1]
\State \textbf{Input:} $\mathbf{F}$, $\mathcal{B}, \mathcal{B'}$, $N_{3}$, $\nu$,  candidate index $n$
\State Compute the SVD of $\mathbf{F}$: $\mathbf{F} = \mathbf{U} \mathbf{\Sigma} \mathbf{V}^H$
\State Extract dominant and sub-dominant right singular vectors:
\Statex \hspace{1.5em} $\mathbf{w}_{\text{opt}} = \mathbf{V}(:,1)$ \quad $\mathbf{w}_{\text{opt}_{\text{sub}}} = \mathbf{V}(:,2)$
\State Select $\mathbf{v}_{l^*,m^*} \in \mathcal{B}$ that maximizes the inner product with the first half of $\mathbf{w}_{\text{opt}}$
\If{$\nu > 1$}
    \State Select $\mathbf{v}_{l^{*'},m^{*'}} \in \mathcal{B'}$ that maximizes the inner product \Statex \hspace{1.5 em}with the first half of $\mathbf{w}_{\text{opt}_{\text{sub}}}$
\EndIf
\For{$t = 1$ to $N_{3}$}
    \State Form the co-phased beam vector:
     $\begin{bmatrix} \mathbf{v}_{l^*,m^*} \\ \varphi_n \mathbf{v}_{l^*,m^*} \end{bmatrix}$\hspace{2.5em}
    \State Determine $n_t$ that maximizes the inner product \Statex \hspace{1.5 em}with~$\mathbf{w}_{\text{opt}}$
\EndFor
\State \textbf{Output:} $\mathbf{v}_{l^*,m^*}$, $\mathbf{v}_{l^{*'},m^{*'}}$ and $\{n_t\}_{t=1}^{N_{3}}$.
\end{algorithmic}
\end{algorithm}

After completing the above selection steps, the final precoder is constructed according to the structure described in Section~II-B, using the selected values of \( \mathbf{v}_{l^*,m^*} \), \( \mathbf{v}_{l^{*'},m^{*'}} \), and \( n_t \). The complete procedure is summarized in Algorithm~\ref{alg:precoder} for the case \( P_{\text{CSI-RS}} < 16 \).
\subsubsection{Time complexity analysis}
The time complexity of the SVD, i.e., $\mathcal{O}(\min(N_T N_R^2, N_R N_T^2))$, where $\mathcal{O}(\cdot)$ denotes the big-O notation, introduces an initial processing overhead for both the proposed and conventional precoder selection methods. During the computation of the wideband part of the precoder in the proposed method (i.e., (\ref{eq:pro_first}) and (\ref{eq:pro_second})), the complexity is \( \mathcal{O}\left(\frac{N_{\mathcal{B}} N_T}{2} \right) \), where \( N_{\mathcal{B}} \) denotes the number of candidate precoding vectors in the codebook \( \mathcal{B} \), for \( \nu = 1 \). Since this step is independent of the number of layers, the additional complexity introduced for \( \nu > 1 \) is only \( \mathcal{O}\left(\frac{N_T}{2} \right) \). As observed in (\ref{eqn:co-phase}), the selection of the co-phasing parameter incurs a time complexity of \( \mathcal{O}(N_3 N_T) \). Consequently, the overall time complexity of the proposed method can be approximately expressed as
\[
\mathcal{O}\left(\min(N_T N_R^2, N_R N_T^2) + N_T\left( \frac{N_{\mathcal{B}}}{2} + N_3 \right) \right).
\] 
If the conventional approach~\cite{tutdlprecoderselection} were used, the computation of the wideband part of the precoder would depend on the number of transmission layers, and the resulting wideband precoding matrix would be used to determine the co-phase parameter by calculating mutual information for each subband. This process would introduce an additional complexity of approximately \( \mathcal{O}(N_T(N_{3} + N_{\mathcal{B}})(\nu - 1)) \) when \( \nu > 1 \). The proposed method eliminates this overhead. This advantage becomes more pronounced in large-scale MIMO systems with numerous transmit antennas and subbands, typical in multi-layer transmission scenarios.
\vspace{-6pt}
\begin{algorithm}[b!]
\caption{MCA-Based RIS Optimization Using Dominant singular values} \label{alg:mca}
\begin{algorithmic}[1]
\State \textbf{Input:} \( \nu \), \( b \), \( T \), \( N_{\text{RIS}} \), \( N_{\text{RIS,new}} \)
\If{\( \nu > 1 \)}
    \State Set the OP metric as the sum of the two largest \Statex \hspace{1.5 em}singular values: \( \lambda_1 + \lambda_2 \)
\ElsIf{\( \nu = 1 \)}
    \State Set the OP metric as the largest singular value: \( \lambda_1 \)
\EndIf
    \State Generate a set \( \Psi = \{ \boldsymbol{\theta}_t \}_{t=1}^T \) of \( T \) random RIS configurations based on \( \Xi_b \)
\For{each \( \boldsymbol{\theta}_t \in \Psi \)}
\State Calculate the OP metric of the channel matrix \( \mathbf{F} \) with \Statex \hspace{1.5 em}RIS configured by \( \boldsymbol{\theta}_t \)
\EndFor
\State Identify the two configurations \( \boldsymbol{\theta}_{\text{max}} \) and \( \boldsymbol{\theta}_{\text{submax}} \) with the highest OP metrics
\State Generate offspring RIS configurations \( \Psi_{\text{son}} \) using Method 1 in \cite{riserank}.
\For{$i = 1, 2, \ldots, 2^{N_{\text{RIS,new}}}$}
    \State Compute the OP metric for \( \mathbf{F} \) configured by offspring \Statex \hspace{1.5 em}\( \boldsymbol{\theta}^{\text{son}}_i \)
\EndFor
\State \textbf{Output:} $\boldsymbol{\theta}_{\text{mca}} = \underset{\boldsymbol{\theta} \in \Psi_{\text{son}}}{\arg\max} \left( {\sum_{r=1}^{\nu} \lambda_r}~/~{\sum_{j=1}^{M} \lambda_j} \right)$.
\end{algorithmic}
\end{algorithm}
\subsection{RIS Optimization Based on singular vectors with MCA}
We introduce a method for the joint optimization of precoding vectors and phase shifts using MCA. We aim to maximize the percentage of first and second singular values. Subsequently, the objective function can be formulated as
\begin{equation}
\label{eq:mca_op}
\begin{aligned}
&\boldsymbol{\theta}_{\text{opt}} = \underset{\boldsymbol{\theta}}{\arg\max}\left(\frac{\sum_{r=1}^{\nu} \lambda_r}{\sum_{j=1}^{M} \lambda_j} \right) \\
&\text{subject to} \quad \theta_n \in \Xi_b,~\forall n = 1, 2, \ldots, N_{\text{RIS}}, \\ 
&~~~~~~~~~~~~~~ \text{and}~M = \min(N_T, N_R).
\end{aligned}
\end{equation}
Here, if $\nu > 1$, we set $\nu = 2$ because the vectors $\mathbf{v}_{l^*,m^*}$ and $\mathbf{v}_{l^{*'},m^{*'}}$, which form the structure in \eqref{eq:W1}, are associated with the first and second dominant singular vectors. Thus, maximizing the corresponding two singular values suffices.

The MCA algorithm can be extended to the following three main steps in scenarios where OP is defined based on either the dominant singular values or the effective rank.
\begin{enumerate}
    \item A total of \( T \) distinct RIS configurations are generated. During each CSI-RS transmission interval, the RIS is sequentially configured using these configurations. For each configuration, the OP is computed. Subsequently, the two configurations yielding the highest OP values are identified and denoted by \( \boldsymbol{\theta}_{\text{max}} \) and \( \boldsymbol{\theta}_{\text{submax}} \), respectively. These configurations are then selected for cross-swapping procedures in the following stages.

    \item $N_{\text{RIS,new}}$ elements of these parent configurations are crossed, as detailed in Method 1\cite{riserank}, to generate $2^{N_{\text{RIS,new}}}$ offspring RIS configurations, denoted $\Psi_{\text{son}} = \{\boldsymbol{\theta}^{\text{son}}_1, \boldsymbol{\theta}^{\text{son}}_2, \ldots, \boldsymbol{\theta}^{\text{son}}_{2^{N_{\text{RIS,new}}}}\}$.

    \item Each $\Psi_{\text{son}}$ offspring RIS configuration is applied sequentially during each CSI-RS transmission. The OP is computed from $\mathbf{F}$, and the final configuration $\boldsymbol{\theta}^{\text{son}}_i$ corresponding to the highest OP is designated as $\boldsymbol{\theta}_{\text{mca}}$.  
    
\end{enumerate}
Algorithm~\ref{alg:mca} formalizes the aforementioned procedures.

\section{Numerical Results}
The entries of the channel matrices $\mathbf{H}$ and $\mathbf{G}$ are generated according to the CDL-C channel model with NLOS scattering, with a delay spread of $300 \times 10^{-9}$ seconds and zero Doppler shift. The carrier frequency is set to $f_c = 4\,\text{GHz}$, corresponding to a wavelength $z = 0.075\,\text{m}$. The OFDM parameters follow the 5G NR specifications with $30\,\text{kHz}$ subcarrier spacing and $2048$ subcarriers. Additional simulation parameters are set as $d_t = 38\,\text{m}$, $d_r = 20\,\text{m}$, $d_x = z/5$, $d_y = z/5$, $G_t = G_r = 1$, $\alpha_t = \alpha_r = 10$, $A_n = 1$, $b = 4$, $T = 200$, and $N_{\text{RIS,new}} = 6$.
\begin{figure}[t]
\centerline{\includegraphics [scale=.60]
{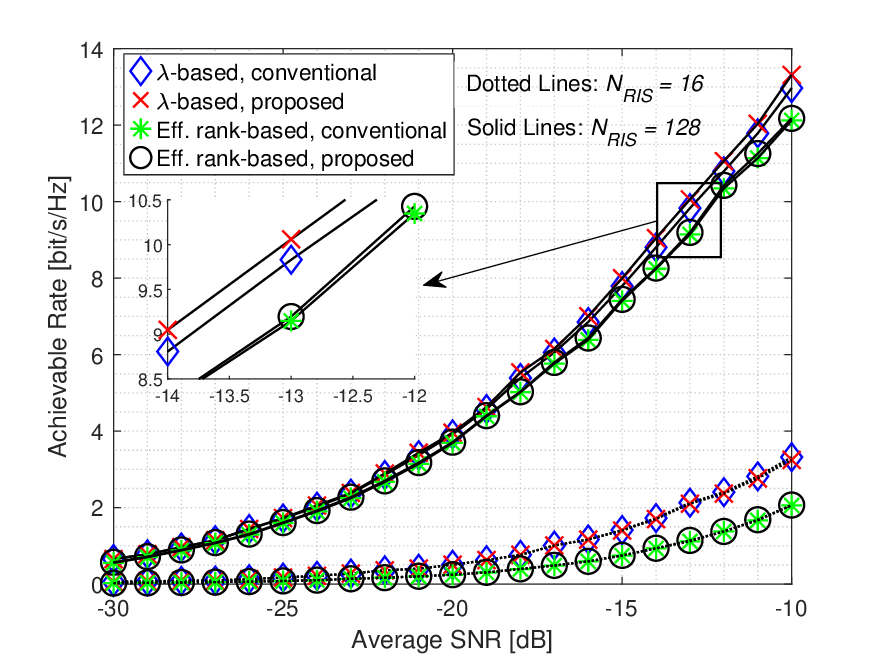}}
\caption{Achievable rate comparison of the proposed and conventional selection methods for different optimization methods and reflective element numbers.}
\label{fig.1}
\end{figure}

Fig.~\ref{fig.1} presents the achievable rate performance of the proposed and conventional precoder selection methods for a 4-layer $4{\times}4$ MIMO system. The results demonstrate that, under $\lambda$-based RIS optimization, the proposed method yields superior performance at higher SNR levels. This improvement is attributed to the SNR gains achieved over the conventional 3GPP scheme—when the number of RIS elements is large—as the method effectively maximizes the dominant singular values of the cascaded channel and leverages the associated singular vectors for precoder selection. In scenarios with fewer RIS elements, the proposed method achieves the same performance as the conventional approach. Furthermore, both selection methods perform similarly under effective rank-based optimization, which yields inferior performance compared to the $\lambda$-based approach. It should be noted, the proposed method maintains a lower complexity than the conventional approach, even when their performance is similar.

\begin{figure}[t]
\centerline{\includegraphics [scale=.60]{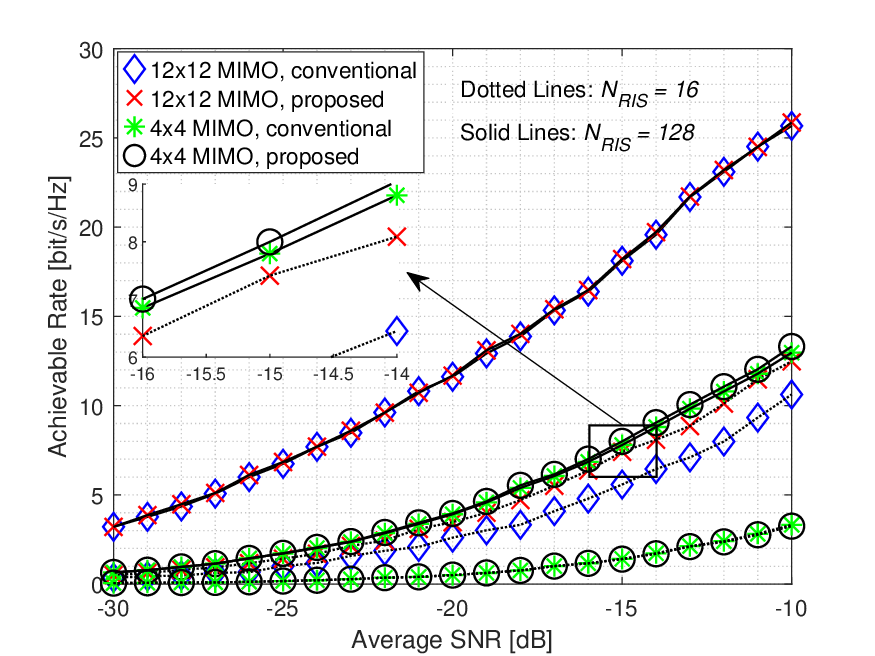}}
\caption{Achievable rates of proposed and conventional selection methods under different antenna configurations for 4-layer and $\lambda$-based~optimization.}
\label{fig.2}
\end{figure}

To evaluate the impact of different MIMO and RIS configurations, Fig.~\ref{fig.2} presents the achievable rate performance under $\lambda$-based RIS optimization for both precoder selection methods. The results indicate that when the RIS has few elements, limited singular value optimization leads to similar performance between the methods in low-dimensional MIMO settings. However, as the number of antennas increases, $\lambda$-based optimization becomes more effective, yielding performance gains for the proposed precoder. On the other hand, when the RIS comprises a large number of elements, $\lambda$-based optimization becomes more effective with a small number of antennas, whereas in larger MIMO systems, distributions of the singular values tend to become more uniform, resulting in similar performance between both methods. 

\section{Conclusion}
In this study, a precoder selection method is proposed for RIS-assisted MIMO systems based on the Type-I codebook defined in 3GPP Releases 15 and 16. The method does not depend on the number of layers when computing the wideband part of the precoder and eliminates the need for mutual information calculations for each subband. RIS phase shifts are optimized using an $\lambda$-based approach that maximizes the dominant singular values of the cascaded channel, and the corresponding singular vectors are used for precoding. To solve this optimization problem, the MCA is applied. The simulation results show that the proposed precoder selection method consistently outperforms the conventional approach under $\lambda$-based RIS optimization, higher gains are observed when the number of antennas and RIS elements is small. In contrast, both methods perform similarly under effective rank-based optimization. Furthermore, in our future research, we plan to implement the proposed precoder selection scheme on SDR-based testbeds within a 5G network, in order to validate its practical feasibility and examine the effects of real-time RIS control and hardware constraints.

\bibliographystyle{IEEEtran}
\bibliography{IEEEabrv,ref}


\end{document}